\documentstyle[12pt]{article}
\newlength{\pubnumber} 

\catcode`\@=11
\@addtoreset{equation}{section}

\def\section{\@startsection{section}{1}{\z@}{3.5ex plus 1ex minus .2ex}
 {2.3ex plus .2ex}{\large\bf}}
\def\subsection{\@startsection{subsection}{2}{\z@}{2.3ex plus .2ex}
 {2.3ex plus .2ex}{\bf}}

\begin{document}

\begin{titlepage}
\samepage{
\setcounter{page}{1}
\vskip 5.5truecm
\begin{center}
  {\large\bf Generator Matrix Elements For $G_2 \supset SU(3)$ II :\\}
  {\large\bf Generic Representations
\footnote{Short title: Generator Matrix Elements For $G_2 \supset
SU(3)$ II.}
\\}

 \vskip 1.5truecm
     {\bf N. Hambli and R. T. Sharp \\}
   {Department of Physics, McGill University\\
    3600 University St., Montr\'eal, Qu\'ebec~H3A-2T8~~Canada}
\footnote{Research supported in part by the Natural Sciences
and Engineering Research Council of Canada and by the Fonds
FCAR du Qu\'ebec.}\\
\vskip 2truecm
\today\\
\end{center}
\vfill\eject
\begin{abstract}
   {Basis states and generator matrix elements are given for the
generic representation $(a,b)$ of $G_2$ in an $SU(3)$ basis.\\}
\end{abstract}
 }
\end{titlepage}

\setcounter{footnote}{0}
\def\beq{\begin{equation}}
\def\eeq{\end{equation}}
\def\beqn{\begin{eqnarray}}
\def\barrl{\begin{array}{ll}}
\def\eeqn{\end{eqnarray}}
\def\earr{\end{array}}


\setcounter{footnote}{0}
\section{ Introduction}

In a recent paper by Farell et al (1994), to be referred
to as I, generator matrix elements of $G_2$
in an $SU(3)$ basis are given for degenerate representation
(one Dynkin label zero) of $G_2$. In this paper the
analysis is extended to generic representations.
\bigskip
\bigskip

In the degenerate case $(a , 0)$ and $(0 , b)$
the subgroup provides a complete set of internal labels,
so the states are automatically orthogonal and can be
normalized straightforwardly; in the generic case
$(a , b)$ there is one missing label and it is
convenient to use non--orthogonal unnormalized states.
\bigskip
\bigskip

Section~2 deals with the basis states and their
non-orthonormality is discussed. In Section~3
the generator matrix elements are derived.
Section~4 contains some concluding remarks.
The earlier paper I contains comments on physical
applications of our results and references to
previous work on the subject; special reference
should be made to the early work of Sviridov,
Smirnov and Tolstoy (1975).
\bigskip
\bigskip

\setcounter{footnote}{0}
\section{ Basis States}

We will use ``character states'', for which the
integrity basis is provided by the $G_2$ character
generator (see I). It is best to start with the
$G_2 \supset SU(3)$ branching rules generating
function (Gaskell et al 1978)
\begin{eqnarray}
G(A,B;P,Q) & = &
\left[(1-AP)\,(1-AQ)\, (1-BP)\, (1-BQ)\right]^{-1}\,\times\,
\nonumber \\
&   &
\left[ (1 - A P Q)^{-1} + B\, (1 - B)^{-1} \right]\, .
\label{ba}
\end{eqnarray}
We are using Dynkin's labelling of the
$G_2$ fundamental representations: $(1 , 0)$ is the
$14-$plet and $(0 , 1)$ the septet. In I and in
Gaskell et al (1978) that numbering is reversed.
\bigskip
\bigskip

The power series expansion of (\ref{ba}) gives
the $G_2 \supset SU(3)$ branching rules:
the coefficient of $A^a\, B^b\, P^p\, Q^q$ is
the multiplicity of the $SU(3)$ representation
$(p , q)$ in the $G_2$ representation $(a , b)$.
But equation (\ref{ba}) does more that count
multiplicities. We interpret $A P\sim \lambda$,
$A Q\sim \nu^*$, $A P Q\sim \alpha$,
$B P\sim \eta$, $B Q\sim \zeta^*$, $B\sim \theta$
as the highest states of the $SU(3)$ representations
contained in the fundamental $G_2$ representations
(see Fig.~1 of I); the highest state of any $SU(3)$
representation in any $G_2$ representation is then
given by the appropriate product of powers of them.
\bigskip
\bigskip

We distinguish two types of state, called
$\alpha$--states and $\theta$--states according
to whether the highest state of the $SU(3)$
representation to which they belong contains
$\alpha$ or $\theta$  (by (\ref{ba}) it cannot
contain both). For $\alpha$--states we adopt
the exponent of $\zeta^*$ as the missing label;
for $\theta$--states the missing label is the
exponent of $\nu^*$. We call the missing label $i$
in both cases. Then a highest $\alpha$--state is
\begin{equation}
\left|
\,{p\;q\;\; i \atop p\; p\; (p + 2q)/3}\,
\right\rangle_{\alpha} =
\eta^{b - i} \; {\zeta^*}^i \; \lambda^{a - q + i} \;
{\nu^*}^{a + b - p - i} \; \alpha^{p + q - a - b} \; ,
\label{bb}
\end{equation}
and a highest $\theta$--state is
\begin{equation}
\left|
\,{p\;q\;\; i \atop p\; p\; (p + 2q)/3}\,
\right\rangle_{\theta} =
\eta^{p - a + i} \; {\zeta^*}^{q - i} \;
\theta^{a + b - p - q} \; \lambda^{a - i} \;
{\nu^*}^i\; .
\label{bc}
\end{equation}
We remark that a state for which $p + q = a + b$ may be
called, and labelled as, an $\alpha-$state or a $\theta-$state
($\alpha$ and $\theta$ both absent); its $i$ label as an
$\alpha-$state is $q$ less its $i$ label as a $\theta-$state.
We suppress the $G_2$ representation
labels $(a , b)$. The internal $SU(3)$ labels
are respectively $t$, $m$, $y$ with the isospin
labels doubled to avoid half--odd values. The ranges
of the labels $p$, $q$, $i$ are such that the
exponents in (\ref{bb}) and (\ref{bc}) take all
non--negative integer values.
\bigskip
\bigskip

States other than the highest ones given by
Equations (\ref{bb}, \ref{bc}) are obtained
by applying the $SU(3)$ lowering generators
$E_{21}$, $E_{32}$, $E_{31}$. In differential
form, suitable for acting on the states
(\ref{bb} , \ref{bc}) each is given by
the sum of the two expressions for it in
Eq.~(I 2.6) and Eq.~(I 2.16).
\bigskip
\bigskip

The basis states of the $G_2$ representation
$(a , b)$ are polynomials of degree a in the
$(1  ,  0)$ states and degree $b$ in the
$(0 , 1)$ states. Thus only stretched
representations (representation labels additive)
in the direct product of $a$ copies of $(1 , 0)$
and $b$ copies of $(0 ,  1)$ are to be retained.
It is known (I) that the elementary unstretched,
and therefore unwanted, representations are all
of degree $2$.
\bigskip
\bigskip

The $G_2$ character generator is needed
in dealing with the unwanted states. Interpreted
as describing the integrity basis for general
basis states, it tells us that certain pairs
of fundamental representation states are
incompatible, {\it i.e.,} never appear
multiplied. The incompatible pairs of a particular
weight are equal in number to the unwanted states
of that weight. Equating each unwanted state to zero
allows us to solve for each incompatible pair in terms
of pairs that are compatible. When an incompatible
pair appears we eliminate it by means of these
incompatibility equations.
\bigskip
\bigskip

We have used a version of the character generator
in which all fundamental basis states appear in
the denominator factors, as opposed to the version
of Gaskell and Sharp (1981) in which only
exterior states appear in denominators. Since
the incompatibility rules characterize the character
generator completely we content ourselves with presenting
the rules in tabular form. Incompatibilities between
$(1 ,  0)$ and $(0 , 1)$ states are shown in
Table~1. Those between pairs of $(1 , 0)$ states are
shown in Figure~2 of I ($\nu$, $\nu^*$, $\delta$ are
compatible with all $(1 , 0)$ states). The only incompatible
pair of $(0 , 1)$ states is $\xi \xi^*$. We should mention
that $\delta$ is the $m = 0$ state of an $SU(2)$ triplet and
$\kappa$ is an $SU(2)$ scalar.
\bigskip
\bigskip

We have determined all the unwanted states and, equating
then to zero, found the equations by which incompatible
pairs are to be eliminated. We give only the replacements
that are actually needed in Section~3.
\bigskip

$(1,0)\times (0,1)$ states
\begin{eqnarray}
\alpha\, \zeta & = &  (2/3)^{1/2} \; \kappa\, \eta -
(4/3)^{1/2} \; \nu^*\, \xi^* -
6^{-1/2} \; \lambda\, \theta \; ,
\nonumber\\
\gamma\, \xi  & = & 2^{-1/2} \; \delta\, \eta +
6^{-1/2} \; \kappa\, \eta - (4/3)^{1/2} \; \nu^*\, \xi^*\; ,
\nonumber\\
\beta\, \xi^*  & = &  -\; \alpha\, \eta^* -
(2/3)^{1/2} \; \kappa\, \zeta^*
-  6^{-1/2} \; \nu^*\, \theta \; ,
\nonumber\\
\alpha\, \theta & = &  (2/3)^{1/2} \; \lambda\, \zeta^*  +
(2/3)^{1/2} \; \nu^*\, \eta \; ,
\nonumber\\
\alpha\, \xi^* & = &  -\; \gamma\, \zeta^*  -
3^{-1/2} \; \lambda\, \eta\;,
\nonumber\\
\lambda^*\, \zeta^* & = & \nu^*\, \eta^* - 2^{-1/2} \; \mu\,
\theta\; ,
\nonumber\\
\mu^* \, \zeta^* & = &  2^{-1/2} \; \lambda\, \theta +
\nu^*\, \xi^*\; ,
\nonumber\\
\lambda\, \zeta & = & \nu\, \eta - 2^{-1/2} \; \mu^*\, \theta\; ,
\nonumber\\
\mu^*\, \xi  & = &  -\; \lambda^*\, \eta - \nu^*\, \zeta \; ,
\nonumber\\
\mu\, \xi^*  & = &  -\; \lambda\, \eta^* - \nu\, \zeta^* \; ,
\nonumber\\
\kappa\, \theta & = &  \nu^*\, \zeta + \nu\, \zeta^* \; ,
\label{bd}
\end{eqnarray}
\bigskip

$(1,0)^2$ states
\begin{eqnarray}
\beta\, \mu^*  & = &  -\; 6^{-1/2} \; \kappa\, \nu^*
- 2^{-1/2} \; \delta\, \nu^*
+ 3^{-1/2} \; \mu\, \lambda \; ,
\nonumber\\
\gamma\, \mu  & = &  1/2 \; \alpha\, \nu +
2^{-1/2} \; \lambda\, \delta +
(12)^{-1/2} \; \mu^*\, \nu^* \; ,
\nonumber\\
\alpha\, \lambda^* & = &  -\; 6^{-1/2} \; \nu^*\, \kappa +
2^{-1/2} \; \nu^*\, \delta - 3^{-1/2}\; \lambda\, \mu \; ,
\nonumber\\
\beta\, \gamma  & = &  6^{-1/2} \; \alpha\, \kappa +
2^{-1/2} \; \alpha\, \delta + 1/3 \; \lambda\, \nu^* \; ,
\nonumber\\
\lambda\, \kappa & = &  (3/2)^{1/2} \; \alpha\, \nu -
2^{-1/2} \; \mu^*\, \nu^* \; .
\nonumber\\
\alpha\, \mu & = & \beta\, \lambda - 3^{-1/2} \; {\nu^*}^2 \; ,
\nonumber\\
\alpha\, \mu^* & = &  3^{-1/2} \; \lambda^2 - \gamma\, \nu^*\; ,
\nonumber\\
\mu\, \mu^* & = &  -\; \nu\, \nu^* - \lambda\, \lambda^*\; ,
\label{be}
\end{eqnarray}
\bigskip

$(0,1)^2$ states
\begin{equation}
\xi\, \xi^* = -\; \eta\, \eta^* - \zeta\, \zeta^*  - 1/2 \; \theta^2\; .
\label{bf}
\end{equation}
It should be remarked that although our states
correspond one--to--one to all states of all
$G_2$ representations they still contain admixtures
of unwanted states belonging to lower representations.
That does not matter for the purpose of computing generator
matrix elements, to which we are about ready to turn.
\bigskip
\bigskip

Since our state are non--orthonormal it is preferable
to define the matrix element
$\left( \, {i} \,\right| \, G\,
\left| \, {j} \, \right)$
of a generator $G$ between two states
$\left| \, {i} \, \right\rangle$
and
$\left| \, {j} \, \right\rangle$
as the coefficient of
$\left| \, {i} \, \right\rangle$
when $G$ acts on
$\left| \, {j} \, \right\rangle$,
rather than as the overlap
$\left\langle \, {i} \,\right| \, G\,
\left| \, {j} \, \right\rangle$.
Matrices so defined can be multiplied in the usual
way. The Wigner--Eckart theorem holds for them. Matrices
for operators can be diagonalized and their eigenvalues
and eigenstates found by the usual standard techniques.
\bigskip
\bigskip

\setcounter{footnote}{0}
\section{ The Generator Matrix Elements}

We now calculate generator matrix elements with respect to
$\alpha-$states (highest state given by (\ref{bb})) and
$\theta-$states (highest state given by (\ref{bc})).
The six significant generators are the components of two
$SU(3)$ tensors $G^{(10)}$ and $G^{(01)}$ which transform
by the indicated $SU(3)$ representations.
\bigskip
\bigskip

According to the $SU(3)$ Wigner--Eckart theorem the matrix
elements of $G^{(10)}$ are given in terms of its reduced
matrix elements (double bars) by
\begin{eqnarray}
 & &  \left\langle
     \,{p_2\;q_2\;\; i' \atop t_2\;m_2\;y_2}\,
   \right| \,G^{(10)}_{t,m,y}\, \left|
     \,{p_1\;q_1\;\; i \atop t_1\;m_1\;y_1}\,
   \right\rangle
     =
   \left\langle \,p_2\;q_2\;\; i'\,
     \left|\!\left| \,G^{(10)}\, \right|\!\right|
   \,p_1\;q_1\;\; i\, \right\rangle
\nonumber \\
& & \times
   \left\langle
     \,{p_1\;q_1 \atop t_1\;m_1\;y_1}\,
      ;
     \,{1\;0 \atop t\;m\;y}\,
   \right| \left.
     \,{p_2\;q_2 \atop t_2\;m_2\;y_2}\,
   \right\rangle
 \times
   \left[(p_2 + 1)(q_2 + 1)(p_2+q_2+2)/2\right]^{-1/2}\; .
\nonumber \\
& &
\label{ca}
\end{eqnarray}
The second factor on the right is an $SU(3)$ Clebsh--Gordon
coefficient. A similar formula exists for the matrix
elements of $G^{(01)}$, we will see below that the
reduced matrix elements of $G^{(01)}$ may be expressed in
terms of those of $G^{(10)}$.
\bigskip
\bigskip

We may write
\begin{eqnarray}
& &
G^{(10)}_{0,0,-{2\over3}}\,
\left| \, {p\;\,q\;\; i \atop p\;p\;(p+2q)/3}\, \right\rangle
= \sum_{i'}\;
\left| \, {p+1\;\,q\;\; i' \atop p\;p\;(p+2q-2)/3}\, \right\rangle \;
A_{i'\;, i} + \nonumber \\
& &
\sum_{i'}\;
\left| \, {p-1\;\,q+1\;\; i' \atop p\;p\;(p+2q-2)/3}\, \right\rangle \;
B_{i'\;, i}\;\; +\;\;
\sum_{i'}\;
\left| \, {p\;\,q-1\;\; i' \atop p\;p\;(p+2q-2)/3}\, \right\rangle \;
C_{i'\;, i}\,\;\; , \nonumber \\
& &
\label{cb}
\end{eqnarray}
where $A_{i'\;, i}$, $B_{i'\;, i}$, $C_{i'\;, i}$ are matrix
elements of $G^{(10)}_{0,0,-{2\over3}}$ to be determined.
We have suppressed a subscript $\alpha$ or $\theta$ on the
states in (\ref{cb}) and, correspondingly, a superscript
$\alpha$ or $\theta$ on the coefficients
$A_{i'\;, i}$, $B_{i'\;, i}$, $C_{i'\;, i}$.
\bigskip
\bigskip

We remark that an ambiguous state $(p + q = a + b)$ is
transformed by $G^{(10)}$ into an $\alpha-$state, an
ambiguous state or a $\theta-$state. An $\alpha-$state
always goes to an $\alpha-$state except when $p + q = a + b + 1$
when it can also go to an ambiguous state; and a
$\theta-$state always goes to a $\theta-$state except
when $p + q = a + b - 1$ when it can also go to an
ambiguous state.
\bigskip
\bigskip

Apply $E_{12}\, E_{23}$ to both side of (\ref{cb}).
The result is
\begin{equation}
G^{(10)}_{1,1,{1\over3}}\,
\left| \, {p\;\,q\;\; i \atop p\;p\;(p + 2q)/3}\, \right\rangle
=
\sqrt{{(p + 1)\, (p + q + 2)}\over{p + 2}}\;
\sum_{i'}\;
\left| \, {p+1\;\,q\;\; i' \atop p\;p\; (p + 2q - 2)/3}\,
\right\rangle \;
A_{i'\;, i}\; .
\label{cc}
\end{equation}
The states in the second and third sums on the right--hand
side of (\ref{cb}) are annihilated and we can read off the
allowed values of $i'$ as well as the matrix element
$A_{i'\;, i}$. We find, for $\alpha-$states, $i'$ can
be only $i$ and
\begin{eqnarray}
{A^{\alpha}_{i\;, i}} & = & (-i - p + a + b)\;
\sqrt{{3\, (2 + p)}\over{(1 + p)\, (2 + p + q)}}\; .
\label{cd}
\end{eqnarray}
For $\theta-$states $i'$ can be $i$ or $i - 1$ and
\begin{eqnarray}
{A^{\theta}_{i\;, i}} & = &  (i - p - q + a + b)\;
\sqrt{{2\, (2 + p)}\over{(1 + p)\, (2 + p + q)}} \; ,
\label{ce}  \\
{A^{\theta}_{i - 1\;, i}} & = & i\; \sqrt{{2\, (2 + p)}\over{(1 + p)\,
(2 + p + q)}} \; .
\label{cf}
\end{eqnarray}
\bigskip
\bigskip

Next apply $E_{23}$ to both sides of $(\ref{cb})$,
after transferring the first sum to the left. The result
is
\begin{eqnarray}
& &
G^{(10)}_{1,-1,{1\over3}}\,
\left| \, {p\;\,q\;\; i \atop p\;p\;(p+2q)/3}\, \right\rangle
\;\; - \nonumber \\
& & \qquad
\sqrt{{p + q + 2}\over{(p + 1)\, (p + 2)}}\;
\sum_{i'}\; A_{i'\;, i}\; E_{21}\;
\left| \, {p+1\;\,q\;\; i' \atop p+1\;p+1\;(p+2q+1)/3}\, \right\rangle \;
 \nonumber \\
& & \qquad \qquad =
\sqrt{q + 1}\; \sum_{i'}\; B_{i'\;, i} \;
\left| \, {p-1\;\,q+1\;\; i' \atop p-1\;p-1\;(p+2q-2)/3}\,
\right\rangle \; .
\label{cg}
\end{eqnarray}
We can read the values of $i'$ and the matrix elements
$B_{i'\;, i}$.
For $\alpha-$states $i'$ can be $i$ or $i + 1$, and
\begin{eqnarray}
{B^{\alpha}_{i\;, i}} & = & {{(i - q + a)\, (2 - i + p + a + b)}
\over{(1 + p)\, \sqrt{1 + q}}} \; ,
\label{ch} \\
{B^{\alpha}_{i + 1\;, i}} & = & {{(i - b)\, (1 - 2 i - p + 2 a + 2
b)}\over{(1 + p)\, \sqrt{1 + q}}} \; .
\label{ci}
\end{eqnarray}
For $\theta-$states $i'$ can also be only $i$ or $i + 1$,
and
\begin{eqnarray}
{B^{\theta}_{i\;, i}} & = & {{(1 + i + p)\,
(-i - p + a)}\over{(1 + p)\, \sqrt{1 + q}}} \; ,
\label{cj} \\
{B^{\theta}_{i + 1\;, i}} & = &  {{(-i + a)\,
(2 + i + p - q + a + b)}\over{(1 + p)\, \sqrt{1 + q}}} \; .
\label{ck}
\end{eqnarray}
Finally transfer the first two sums to the left side of
(\ref{cb}). Only the third sum remains on the right
and we can read the allowed values of $i'$ and the matrix
elements $C_{i'\;, i}$ :
For $\alpha-$states $i'$ can be $i$, $i - 1$ or $i + 1$ and
\begin{eqnarray}
{C^{\alpha}_{i\;, i}} & = &
{1\over{\sqrt{3}\, (1 + p)\, (1 + q)\, (2 + p + q)}}\;
(-4 i - 4 i^2 + 4 i^3 - 6 p -
\nonumber \\
& & \;\;
7 i p - 2 i^2 p + 3 i^3 p - 8 p^2 - 4 i p^2 + i^2 p^2 - 2 p^3 - i p^3 -
\nonumber \\
& & \;\;
6 q - 3 i q - 4 i^2 q + i^3 q - 13 p q - 7 i p q -
2 i^2 p q - 8 p^2 q -
\nonumber \\
& & \;\;
3 i p^2 q - p^3 q - 5 q^2 -
i q^2 - i^2 q^2 - 7 p q^2 - 2 i p q^2 - 2 p^2 q^2 -
\nonumber \\
& & \;\;
q^3 - p q^3 + 6 b + 2 i b - 8 i^2 b + 6 p b - 6 i^2 p b -
i p^2 b +
\nonumber \\
& & \;\;
2 q b + 6 i q b - 2 i^2 q b + 4 p q b + 3 i p q b +
p^2 q b + 2 i q^2 b + p q^2 b +
\nonumber \\
& & \;\;
2 b^2 + 4 i b^2 + 2 p b^2 +
3 i p b^2 - 2 q b^2 + i q b^2 - p q b^2 - q^2 b^2 +
\nonumber \\
& & \;\;
6 a - 3 i a -
4 i^2 a + 3 p a - 3 i p a - 3 i^2 p a - 4 p^2 a - p^3 a +
2 i q a -
\nonumber \\
& & \;\;
i^2 q a -
2 p q a + i p q a - 2 p^2 q a -
q^2 a + i q^2 a - p q^2 a + 7 b a + 3 i b a +
\nonumber \\
& & \;\;
7 p b a + 2 i p b a - q b a + i q b a - q^2 b a + b^2 a + p b^2 a +
 5 a^2 -
\nonumber \\
& & \;\;
i a^2 + 5 p a^2 - i p a^2 + q a^2 + p q a^2 + 2 b a^2 +
 2 p b a^2 + a^3 + p a^3 ) ,
\label{cl} \\
{C^{\alpha}_{i + 1\;, i}} & = & {{(i - b)\,
(-i - p + a + b)}\over{\sqrt{3}\, (1 + p)\, (1 + q)\, (2 + p + q)}}\;
( 1 - 3 i - p - 2 i p - p^2 - i q -
\nonumber \\
& &
\;\;
 p q + 3 b + 2 p b + q b + 3 a + 2 p a + q a ) \; ,
 \label{cm} \\
{C^{\alpha}_{i - 1\;, i}} & = & {{i\, (-4 + i - p - 2 q - a - b)\,
(1 + i + a)}\over{\sqrt{3}\, (1 + q)\, (2 + p + q)}} \; .
\label{cn}
\end{eqnarray}
For $\theta-$states $i'$ takes the values $i$ or $i - 1$ and
\begin{eqnarray}
{C^{\theta}_{i\;, i}} & = & {{(i - q)\, (1 - i + q + a)\,
 (4 + i + p + q + a + b)}\over{\sqrt{2}\, (1 + q)\, (2 + p + q)}} \; ,
\label{co} \\
{C^{\theta}_{i - 1\;, i}} & = & {{i\, (1 + i + p)\,
 (-i - p + a)}\over{\sqrt{2}\, (1 + q)\, (2 + p + q)}} \; .
\label{cp}
\end{eqnarray}
When one of the states in the matrix element is ambiguous
its label $(\alpha)$ or $(\theta)$ must be the same as for
the other state. If both are ambiguous they should both be
given the same label $(\alpha)$ or $(\theta)$.
\bigskip
\bigskip

The matrix elements
$A^{(\alpha\,,\,\theta)}_{i'\;, i}$,
$B^{(\alpha\,,\,\theta)}_{i'\;, i}$,
$C^{(\alpha\,,\,\theta)}_{i'\;, i}$ are all we need
to obtain the corresponding reduced matrix elements.
Also we can get the needed matrix elements of $G^{(01)}$
by following the steps used above for $G^{(10)}$.
If we use lowest states (and near lowest) of the
$SU(3)$ IR $(q , p)$ instead of highest states
(and near highest) of $(p , q)$, and use hermitian
conjugates of all the generators, then apart from some
changes in phase, the steps are identical in every detail,
with starred and unstarred variables interchanged.
\bigskip
\bigskip

For $\alpha-$states the results are
\begin{eqnarray}
& &
   \left\langle \,p+1\;q\;i\,
       \left|\!\left| \,G^{(10)}\, \right|\!\right|
   \,p\;q\;i\,\right\rangle
        =
       -\; \left\langle \,q\;p+1\;i\,
       \left|\!\left| \,G^{(01)}\, \right|\!\right|
   \,q\;p\;i\, \right\rangle
         \nonumber \\
& &  = (-i - p + a + b)\;
\sqrt{{3\, (2 + p)\, (1 + q)\, (3 + p + q)}\over 2}\; ,
\label{cq}
\end{eqnarray}
\begin{eqnarray}
& &
   \left\langle \,p-1\;q+1\;i\,
      \left|\!\left| \,G^{(10)}\, \right|\!\right|
   \,p\;q\; i\, \right\rangle
       =
      \left\langle \,q+1\;p-1\;i\,
      \left|\!\left| \,G^{(01)}\, \right|\!\right|
   \,q\;p\;i\, \right\rangle
        \nonumber \\
& &  = {(i - q + a)\, (2 - i + p + a + b)}\;
\sqrt{{(2 + q)\, (2 + p + q)}\over{2\, (1 + p)}}\; ,
\label{cr}
\end{eqnarray}
\begin{eqnarray}
& &
   \left\langle \,p-1\;q+1\;i+1\,
      \left|\!\left| \,G^{(10)}\, \right|\!\right|
   \,p\;q\; i\, \right\rangle
       =
      \left\langle \,q+1\;p-1\;i+1\,
      \left|\!\left| \,G^{(01)}\, \right|\!\right|
   \,q\;p\;i\, \right\rangle
        \nonumber \\
& & = {(i - b)\, (1 - 2 i - p + 2 a + 2 b)}\;
\sqrt{{(2 + q)\, (2 + p + q)}\over{2\, (1 + p)}}\; ,
\label{cs}
\end{eqnarray}
\begin{eqnarray}
& &
   \left\langle \,p\;q-1\;i\,
      \left|\!\left| \,G^{(10)}\, \right|\!\right|
        \,p\;q\;i\, \right\rangle
       =
   -\;\left\langle \,q-1\;p\;i\,
      \left|\!\left| \,G^{(01)}\, \right|\!\right|
   \,q\;p\;i\, \right\rangle
         \nonumber \\
& & =
{1\over\sqrt{{6\, (1 + p)\, (1 + q)\, (2 + p + q)}}}\;
(-4 i - 4 i^2 + 4 i^3 - 6 p -
\nonumber \\
& & \;\;
7 i p - 2 i^2 p + 3 i^3 p - 8 p^2 - 4 i p^2 + i^2 p^2 - 2 p^3 - i p^3 -
\nonumber \\
& & \;\;
6 q - 3 i q - 4 i^2 q + i^3 q - 13 p q - 7 i p q -
2 i^2 p q - 8 p^2 q -
\nonumber \\
& & \;\;
3 i p^2 q - p^3 q - 5 q^2 -
i q^2 - i^2 q^2 - 7 p q^2 - 2 i p q^2 - 2 p^2 q^2 -
\nonumber \\
& & \;\;
q^3 - p q^3 + 6 b + 2 i b - 8 i^2 b + 6 p b - 6 i^2 p b -
i p^2 b +
\nonumber \\
& & \;\;
2 q b + 6 i q b - 2 i^2 q b + 4 p q b + 3 i p q b +
p^2 q b + 2 i q^2 b + p q^2 b +
\nonumber \\
& & \;\;
2 b^2 + 4 i b^2 + 2 p b^2 +
3 i p b^2 - 2 q b^2 + i q b^2 - p q b^2 - q^2 b^2 +
\nonumber \\
& & \;\;
6 a - 3 i a -
4 i^2 a + 3 p a - 3 i p a - 3 i^2 p a - 4 p^2 a - p^3 a +
2 i q a -
\nonumber \\
& & \;\;
i^2 q a -
2 p q a + i p q a - 2 p^2 q a -
q^2 a + i q^2 a - p q^2 a + 7 b a + 3 i b a +
\nonumber \\
& & \;\;
7 p b a + 2 i p b a - q b a + i q b a - q^2 b a + b^2 a + p b^2 a +
 5 a^2 -
\nonumber \\
& & \;\;
i a^2 + 5 p a^2 - i p a^2 + q a^2 + p q a^2 + 2 b a^2 +
2 p b a^2 + a^3 + p a^3 ) \; ,
\label{ct}
\end{eqnarray}
    \begin{eqnarray}
& &
   \left\langle \,p\;q-1\;i+1\,
      \left|\!\left| \,G^{(10)}\, \right|\!\right|
        \,p\;q\;i\, \right\rangle
       =
   -\;\left\langle \,q-1\;p\;i+1\,
      \left|\!\left| \,G^{(01)}\, \right|\!\right|
   \,q\;p\;i\, \right\rangle
         \nonumber \\
& & = {{(i - b)\,
(-i - p + a + b)}\over{\sqrt{6\, (1 + p)\, (1 + q)\, (2 + p + q)}}}\;
( 1 - 3 i - p - 2 i p - p^2 - i q -
\nonumber \\
& &
\;\;
 p q + 3 b + 2 p b + q b + 3 a + 2 p a + q a ) \; ,
\label{cu}
\end{eqnarray}
\begin{eqnarray}
& &
   \left\langle \,p\;q-1\;i-1\,
      \left|\!\left| \,G^{(10)}\, \right|\!\right|
        \,p\;q\;i\, \right\rangle
       =
   -\;\left\langle \,q-1\;p\;i-1\,
      \left|\!\left| \,G^{(01)}\, \right|\!\right|
   \,q\;p\;i\, \right\rangle
         \nonumber \\
& & = {i\, (-4 + i - p - 2 q - a - b)\,
(1 + i + a)}\;
\sqrt{{1 + p}\over{6\, (1 + q)\, (2 + p + q)}}\; .
\label{cv}
\end{eqnarray}
For $\theta-$states we find for the reduced matrix
elements
\begin{eqnarray}
& &
   \left\langle \,p+1\;q\;i\,
       \left|\!\left| \,G^{(10)}\, \right|\!\right|
   \,p\;q\;i\,\right\rangle
        =
       -\; \left\langle \,q\;p+1\;i\,
       \left|\!\left| \,G^{(01)}\, \right|\!\right|
   \,q\;p\;i\, \right\rangle
         \nonumber \\
& & = (i - p - q + a + b)\;
\sqrt{3\, (2 + p)\, (1 + q)\, (3 + p + q)}\; ,
          \label{cw}
\end{eqnarray}
\begin{eqnarray}
& &
   \left\langle \,p+1\;q\;i-1\,
       \left|\!\left| \,G^{(10)}\, \right|\!\right|
   \,p\;q\;i\,\right\rangle
        =
      -\; \left\langle \,q\;p+1\;i-1\,
       \left|\!\left| \,G^{(01)}\, \right|\!\right|
   \,q\;p\;i\, \right\rangle
         \nonumber \\
& & = i\; \sqrt{{3\, (2 + p)\, (1 + q)\,
(3 + p + q)}\over 2}\; ,
\label{cx}
\end{eqnarray}
\begin{eqnarray}
& &
   \left\langle \,p-1\;q+1\;i\,
      \left|\!\left| \,G^{(10)}\, \right|\!\right|
   \,p\;q\;i\, \right\rangle
       =
      \left\langle \,q+1\;p-1\;i\,
      \left|\!\left| \,G^{(01)}\, \right|\!\right|
   \,q\;p\;i\, \right\rangle
        \nonumber \\
& & = {(1 + i + p)\, (-i - p + a)}\;
\sqrt{{(2 + q)\, (2 + p + q)}\over{2\, (1 + p)}}\; ,
          \label{cy}
\end{eqnarray}
\begin{eqnarray}
& &
   \left\langle \,p-1\;q+1\;i+1\,
      \left|\!\left| \,G^{(10)}\, \right|\!\right|
   \,p\;q\; i\, \right\rangle
       =
      \left\langle \,q+1\;p-1\;i+1\,
      \left|\!\left| \,G^{(01)}\, \right|\!\right|
   \,q\;p\;i\, \right\rangle
        \nonumber \\
& & = {(-i + a)\, (2 + i + p - q + a + b)}\;
\sqrt{{(2 + q)\, (2 + p + q)}\over{2\, (1 + p)}}\; ,
          \label{cz}
\end{eqnarray}
\begin{eqnarray}
& &
   \left\langle \,p\;q-1\;i\,
      \left|\!\left| \,G^{(10)}\, \right|\!\right|
        \,p\;q\;i\, \right\rangle
       =
   -\;\left\langle \,q-1\;p\;i\,
      \left|\!\left| \,G^{(01)}\, \right|\!\right|
   \,q\;p\;i\, \right\rangle
         \nonumber \\
& & = {(i - q)\, (1 - i + q + a)\, (4 + i + p + q + a + b)}\;
\sqrt{{1 + p}\over{4\, (1 + q)\, (2 + p + q)}}\; ,
\nonumber \\
& &
\label{cza}
\end{eqnarray}
\begin{eqnarray}
& &
   \left\langle \,p\;q-1\;i-1\,
      \left|\!\left| \,G^{(10)}\, \right|\!\right|
        \,p\;q\;i\, \right\rangle
       =
   -\;\left\langle \,q-1\;p\;i-1\,
      \left|\!\left| \,G^{(01)}\, \right|\!\right|
   \,q\;p\;i\, \right\rangle
         \nonumber \\
& &  =  {i\, (1 + i + p)\, (-i - p + a)}\;
\sqrt{{1 + p}\over{4\, (1 + q)\, (2 + p + q)}}\; .
\label{czb}
\end{eqnarray}
\bigskip
\bigskip

\section{ Concluding Remarks}

We have used non-orthonormal generic states; that is
convenient when there is a missing label. Our results
can be compared with those of I by setting one of the $G_2$
representation $a$ or $b$ labels equal to zero,
omitting the ``missing'' label $i$, and replacing the
normalization constants $N_{p,q}$ of I by unity.

\input epsf.tex
\begin{table}[p]
\centerline{\epsfxsize 5 truein \epsfbox {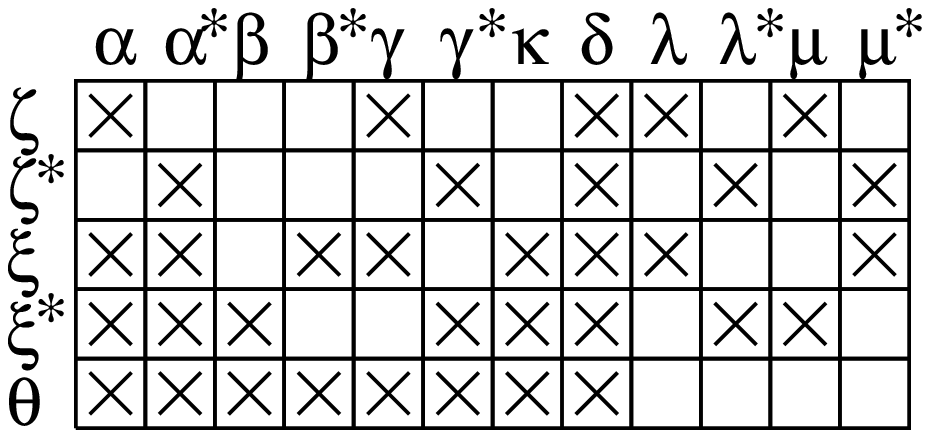}}
\vskip -3.83 truein
\caption{Incompatibility table. Each incompatible pair of
$(1 , 0)\times (0 , 1)$ states is marked with a cross.
The $(0 , 1)$ states $\eta$, $\eta^*$ and the $(1 , 0)$
states $\nu$, $\nu^*$ are compatible with all.}
\end{table}

\pagebreak

\setcounter{footnote}{0}
\section{ References}

Farell L, Lam C S and Sharp R T
1994 $G_2$ generator matrix elements for degenerate
representations in an $SU(3)$ basis.
{\it J. Phys. A: Math. Gen.} {\bf 27} 2761 -- 2771;
we refer to this paper as I.
\medskip

Gaskell R, Peccia A and Sharp R T
1978 Generating functions for polynomial irreducible tensors.
{\it J. Math. Phys.} {\bf 19} 727 -- 733.
\medskip

Gaskell R and Sharp R T
Generating functions for $G_2$ characters and
subgroup branching rules.
1981 {\it J. Math. Phys.} {\bf 22} 2736 -- 2739.
\medskip

Sviridov D T, Smirnov Y F and Tolstoy V N 1975
On the structure of the irreducible representation
basis for the exceptional group $G_2$.
{\it Rep. Math. Phys.} {\bf 7} 349 -- 360.
\end{document}